# Mobile Crowd Sensing and Computing: When Participatory Sensing Meets Participatory Social Media

Bin Guo[1], Chao Chen[2,3], Daqing Zhang[3], Zhiwen Yu[1], Alvin Chin[4]

[1] School of Computer Science, Northwestern Polytechnical University, P. R. China
[2] Chongqing University, P. R. China
[3] Institut TELECOM SudParis, France
[4] Microsoft
guob@nwpu.edu.cn

**Abstract**

With the development of mobile sensing and mobile social networking techniques, Mobile Crowd Sensing and Computing (MCSC), which leverages heterogeneous crowdsourced data for large-scale sensing, has become a leading paradigm. Built on top of the participatory sensing vision, MCSC has two characterizing features: (1) it leverages heterogeneous crowdsourced data from two data sources: participatory sensing and participatory social media; and (2) it presents the fusion of human and machine intelligence (HMI) in both the sensing and computing process. This paper characterizes the unique features and challenges of MCSC. We further present early efforts on MCSC to demonstrate the benefits of aggregating heterogeneous crowdsourced data.

## 1. Introduction

The effective use of the incredible and continuous production of data coming from different sources (e.g., enterprises, Internet of Things, online systems, etc.) will transform our life and work. Within this context, people are not only data consumers, but participate in different ways (e.g., smartphone sensing, online posting) in the data production process. In this article, we discuss the opportunities that heterogeneous human participation offer to systems and services that rely on large-scale sensing.

It is essential to firstly clarify the motivation of taking human-in-the-loop for large-scale sensing. In the past few years, researchers have studied the benefits of understanding urban/community dynamics [1]. However, traditional stationary wireless sensor network deployments often fail to capture such dynamics either because they do not have enough sensing capabilities or are limited in terms of scalability (e.g., high deployment and maintenance cost). Mobile Crowd Sensing and Computing (MCSC), nevertheless, offers a new way for large-scale sensing and computing. On one hand, the sheer number of mobile devices (e.g., smartphones, tablets, wearable devices) and their inherent mobility provide the ability to sense and infer people's context (e.g., ambient noise) in an unprecedented manner. On the other hand, highly-scalable sensing with mobile devices in combination with cloud computing support gives MCSC systems those scalability and versatility properties that often are lacking in static deployments. Although it is quite difficult to attempt a formal definition of the MCSC paradigm, we could state that MCSC is *a new sensing paradigm that empowers ordinary people to contribute data sensed or generated from their mobile devices, aggregates and processes heterogeneous crowdsourced data in the cloud for intelligent service provision*.

From the AI perspective, MCSC is founded on a distributed problem-solving model where crowds are engaged in complex problem solving procedures through open calls. The concept of crowd-powered problem solving has been explored in several research areas. The term "*crowdsourcing*" was coined in 2005 by the Wired Magazine. The definition of the term crowdsourcing is as follows[1]: *the practice of obtaining needed services or content by soliciting contributions from a large group of people, and especially from an online community*. Wikipedia[2], where thousands of contributors from across the globe have collectively created the world's largest encyclopedia, is a typical example. MCSC extends this concept by going beyond the boundaries of online communities and reaching out to the mobile device user population for sensing participation. With *participatory sensing*, first proposed by Burke et al. [2], we see for the first time solutions that require explicit people involvement in accomplishing sensing tasks. MCSC broadens the concept of participatory sensing from two aspects. First, it takes advantages of various forms of human participation in the mobile Internet era. Generally speaking, MCSC sensing modalities can be obtained from specific hardware sensors (e.g., accelerometers, cameras) available on mobile devices and from the information trail (e.g., social media posts) directly generated by users. Second, MCSC presents the fusion of human and machine intelligence in both the sensing and computing process. The usage of heterogeneous crowdsourced data as well as the integration of human and machine intelligence opens up new and unexpected opportunities. We use the following trip planning scenario to showcase the characters of MCSC.

---

[1] Source: Merriam-Webster Dictionary, http://www.merriam-webster.com/dictionary/crowdsourcing

[2] http://en.wikipedia.org/

*Trip planning is a typical MCSC application. With participatory sensing, we can collect GPS trajectory data from vehicles and compute the optimal route when answering a query with departure and destination points. However, for a more complex query, that is, to generate an itinerary for a visitor to a city given the time budget (start time, end time), it is not possible to leverage a single trajectory dataset. Further information such as point of interests (POIs) in the city, the best time to visit the POIs, and user preferences to different POIs, are further needed. These information can be obtained, however, by reusing the user-generated data from location-based social networks (LBSNs).*

The above scenario demonstrates the aggregated power of participatory sensing and social networks for intelligent service provision. The key contributions of this paper can be summarized as follows:

- Characterizing the main features of MCSC by combining participatory sensing and participatory social media.
- Exploring the fusion of human and machine intelligence in MCSC, and discussing about the key research challenges such as cross-space data mining and data quality maintenance.
- Presenting several representative studies to demonstrate the power and usage of MCSC, including two of our recent works and the ones from other research groups.

## 2. Merging Participatory Sensing and Participatory Social Media

MCSC combines two distinct data generation modes: *participatory sensing* advocates the involvement of citizens in the sensing loop in the real world; *participatory social media* refers to user-generated data in mobile social networks (MSNs). MSN services can bridge the gap between online interaction and physical elements (e.g., check-in places, activities), and the data collected from them provides another way to understand urban dynamics.

### 2.1. User Participation: Explicit or Implicit

Previous studies often discuss about user-participated data sourcing in one dimension: *the degree of user involvement in the sensing process*. As presented by Ganti et al. [3], crowd-powered sensing can span a wide spectrum in terms of user involvement, with participatory and opportunistic sensing at the two ends. The proportion of human involvement depends on application requirements and device capabilities. With the two data generation modes in MCSC, we intend to categorize the sensing style from a new dimension: *the data usage manner*. For both participatory and opportunistic sensing, data collection is the primary purpose of the application. The sensing task is therefore *explicit* to its participants. For user-generated data in MSNs, however, the data is used for a second purpose (the primary purpose is social interaction) and is often *implicit* to the contributors. We thus categorize the sensing style of MCSC into two basic types regarding to data usage: *explicit* and *implicit*. Note that sometimes the two basic sensing styles can be merged as a complex one. For instance, explicit crowdsensed data can also be used for a second purpose (e.g., using crowdsensed urban-noise data to predict the functions of regions in a city), thus integrating explicit sensing with implicit sensing.

## 2.2. Data Control for Primary/Secondary Use

The sharing of personal data in MCSC applications can raise significant privacy concerns, with information (e.g., locations, preferences) being sensitive and vulnerable to privacy attacks. To motivate user participation, we should explore new techniques that protect user privacy. We identify two basic sensing styles (*explicit* and *implicit*) for MCSC, and they have distinct data control needs.

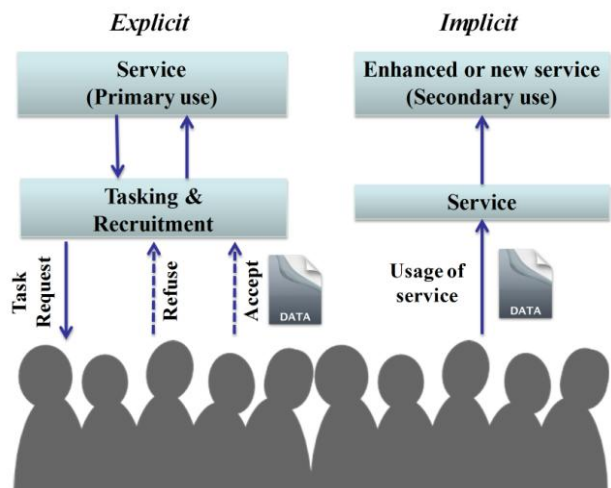

Fig. 1. Data control in MCSC

In *explicit sensing*, sensor data collection is triggered by tasks, which specify the sensing modalities (e.g., regions of interest, sampling context) based on application requests. The tasks are distributed to the mobile device carriers that satisfy the tasking requirements, and people can decide to accept or refuse the task allocated (as shown in Fig. 1). We can find that data is collected under the "primary-use" manner in explicit sensing. Privacy in explicit sensing should guarantee that participants maintain control over the release of their sensitive information, e.g., the degree of granularity and data recipients.

In *implicit sensing*, data is contributed not for a sensing task, but for users to enjoy online services (socializing in Facebook, purchasing goods in Amazon). As shown in Fig. 1, the data is reused to enhance original services or create new services by third-parties (i.e., used for a second purpose). Since most innovative secondary uses are not imagined when the data is collected, how should we protect user privacy in implicit sensing?

"Terms & Conditions" takes an important role in current online services, where people are told at the time of service usage which information is being gathered and for what purpose. This law works for explicit sensing, but, as discussed above, may not work well for implicit sensing. For example, it is often difficult to pre-specify the (secondary) purpose. Improved data-notice strategies should thus be studied. One basic rule to follow is that we should have users keep sense of awareness (and control) to any use of their data. We may need to build an "*evolving notice center (ENC)*" for each service, to make users aware what data is collected, how long will the data be kept, and who are using their data. Enhanced legal protections to the service providers on data reuse can also alleviate user concerns in implicit sensing.

## 2.3. Fusion of Human and Machine Intelligence

The primary feature of MCSC is having various human participation (e.g., locating sensing objects, capturing pictures, posting in MSNs) in the large-scale problem solving process. The coexistence of human and machine power, however, needs to be orchestrated in an optimal manner to enhance them both. An important reason to combine human and machine intelligence is that they often show distinct strengths and weaknesses, as illustrated in Fig. 2. We term the fusion of human intelligence (HI) and machine intelligence (MI) as HMI, which *characterizes the complementary roles of HI and MI in problem solving and integrates them for MCSC service provision*.

As shown in Fig. 2, there are three important layers in a generic MCSC framework, and HI and MI work collaboratively over all these layers. For example, in the *crowd sensing* layer, machines can allocate tasks to proper participants according to the task needs and human behavior patterns, and the selected workers can execute the assigned tasks using their cognition/perception abilities. In the *data transmission* layer, human mobility patterns and social interactions facilitate the development of optimized networking methods [4]. In the *data processing* layer, the integrated power of human and machine intelligence can attain higher performance (e.g., accuracy on classification) than either of them.

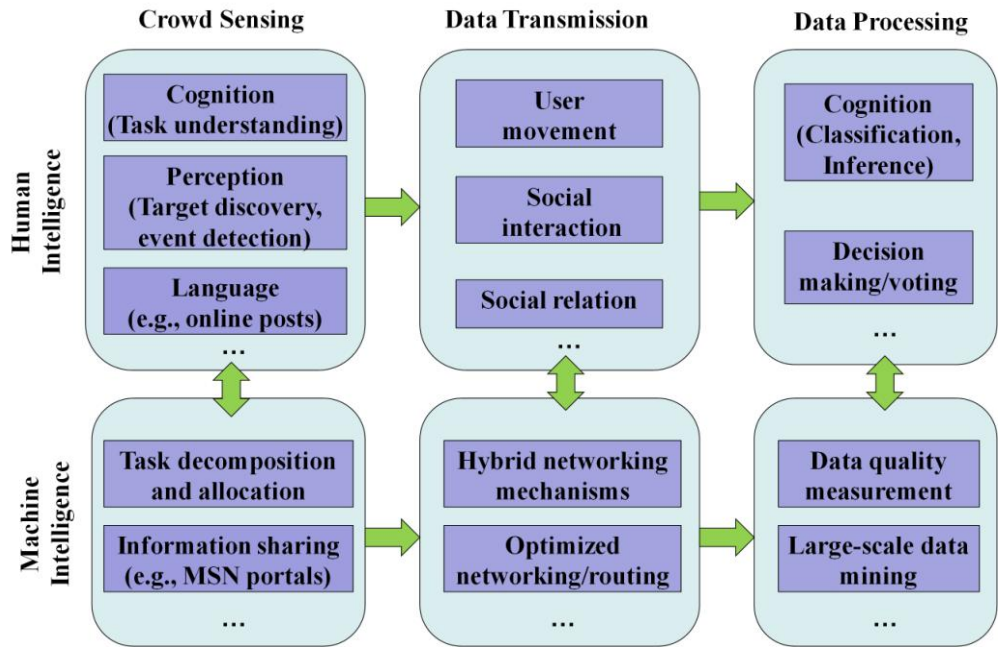

Fig. 2. The Fusion of Human and Machine Intelligence in MCSC

## 3. Key Research Challenges

The combination of two participatory data-generation modes in MCSC also raises new research challenges and issues, some of which are discussed below.

### 3.1. Heterogeneous, Cross-Space Data Mining

The strength of MCSC relies on the usage of crowdsourced data from both physical and virtual societies. For the same sensing object (e.g., a social gathering in a street corner), it will interact with both spaces and leave fragmented data in each space, making the information obtained from different communities (online or offline) different. For instance, we can learn social relationships from online social networks, and infer group activities using smartphone sensing in the real world. Obviously, the complementary nature of heterogeneous communities will bring new opportunities to develop new human-centric services. Therefore, we should integrate and fuse the information from heterogeneous, cross-space data sources — we term it as *cross-space data mining* — to attain a comprehensive picture about the sensing object. Potential research issues include how to reveal the complex linkage and interplay among the data from online/offline space, how to aggregate the information from heterogeneous data sources for enhanced learning, and so on. Recent progress on this has been discussed in detail in our previous work [5].

### 3.2. Potential Fusion Patterns of HMI

The successful of MCSC largely depends on the fusion of human and machine intelligence. However, the challenge is how HI and MI should be fused to produce aggregated effects. In other words, we should study the potential fusion patterns of HI and MI to attain HMI. Three potential patterns are identified in this article, namely *sequential*, *parallel*, and *hybrid*. The *sequential* pattern is often used in MCSC. For example, given a crowd sensing task, machines can decompose it into sub-tasks and people can execute the assigned sub-tasks using their cognition abilities. HI and MI can also be combined in a *parallel* manner. Still taking the accomplishment of a complex sensing task for example, human nodes and machines (e.g., static sensing nodes) may have complementary sensing abilities, and need to work in a parallel way to fully capture the required information. Finally, two or more parallel or sequential units can be integrated in a *hybrid* manner when more complex problems are to be solved.

### 3.3. Data Quality and Selection

The involvement of human participation in the sensing process also brings forth certain uncertainties to MCSC systems. For example, anonymous participants may send incorrect, low-quality, or even fake data to a data center [6]. Data contributed by different people may be redundant or inconsistent. Certain quality estimation and prediction methods are thus necessary to evaluate the quality of sensing data, and statistical processing can be used to identify outliers. Data selection is also crucial to filter low-quality or irrelevant data and generate a high-quality dataset for further data processing or information presentation. For instance, Ding et al. [6] proposed a data cleansing-based robust spectrum sensing algorithm to eliminate the negative impact of abnormal or low-quality data in crowd sensing. SmartPhoto [7] quantified the utility of crowdsourced photos based on the associated contextual information, such as the smartphone's orientation, position, and location.

# 4. MCSC on the Road

The study of MCSC brings new potential in many application areas. This section firstly makes a summary of two of our ongoing works. The first work is a trip planning application that demonstrates the power of using a combination of participatory sensing and social media data. The second work illustrates our efforts on the fusion of human and machine intelligence in MCSC. We also use more examples from other research groups to demonstrate the power of MCSC.

## 4.1. Trip Planning with Heterogeneous Crowdsourced Data

In the introduction, we have described the trip planning scenario. A detailed analysis of the problem as well as our solution is presented below. As shown in Fig. 3, in order to plan a trip for visiting a popular tourist city, one needs to select a number of user-preferred POIs among hundreds of available venues ($V_1$ to $V_5$ in Fig. 3), figure out the order in which they are visited, ensure the total time it takes to visit them (the *stay time*) and transit from one venue to the next (the *transit time*), and meet the user's *time budget*.

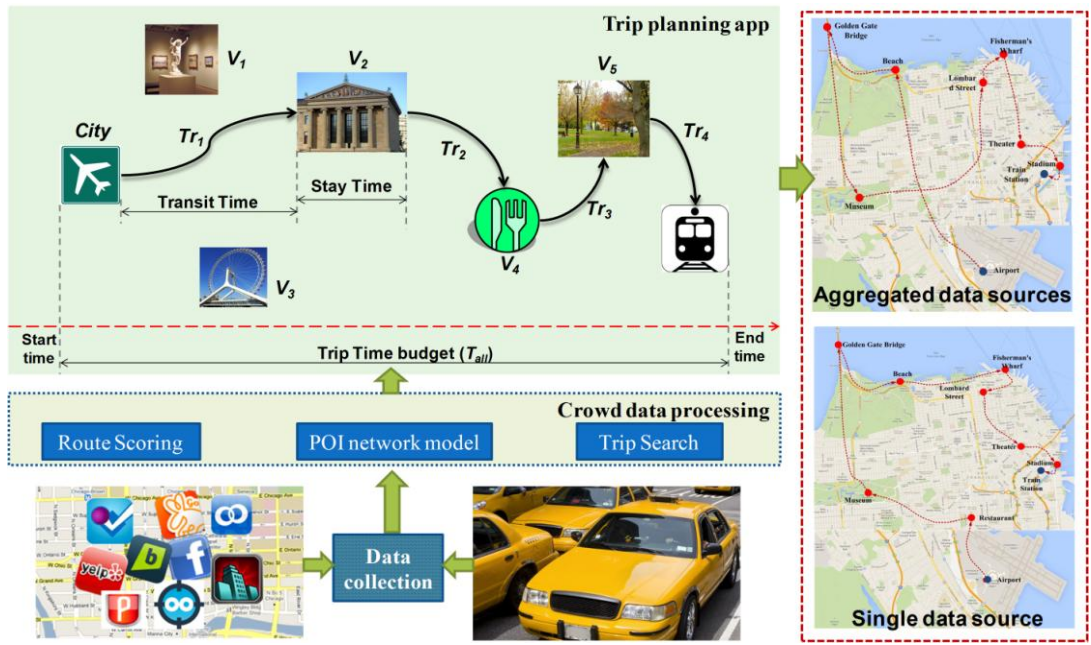

Fig. 3: Trip planning over heterogeneous crowdsourced data

In order to address the trip planning problem, the information about the POIs and links among POIs need to be acquired to build a POI network model. Two types of crowdsourced data sources can be exploited: *i)* LBSNs (e.g., FourSquare[1]), which can inform the popularity (the average number of visitors per day), operating hours, and the best visiting time of the POIs, and an individual user's visiting history; and *ii)* GPS trajectories of people and taxis, which can indicate the stay time in each place and the transit time between two places. Previous studies rely on either of the two data sources, which results in incomplete POI network models. For instance, Cao et al. [8] assumed that the transit time between any two POIs is static and proportionate to their distance. It does not work in real situations because the transit time between venues can be significantly distinct at different timeslots due to dynamic traffic conditions.

In view of the above reasons, we propose a personalized, traffic-aware trip planning framework called TripPlanner [9], which leverages heterogeneous crowdsourced digital footprints for POI network model construction. More specifically, we make use of two data sources including location data sensed by mobile vehicles (taxi GPS trace data) and user-generated data in FourSquare considering their complementary nature. The TripPlanner system consists of two major components: *route scoring* and *trip search* (see Fig. 3). The route scoring module is responsible for estimating the attractiveness of a candidate route to a given user, where two factors are considered: (a) the attractiveness of a venue to the user, (b) the suitability of the visiting time to each venue. The trip search module applies heuristic algorithms to add user-preferred venues iteratively to the generated routes, with the objective of maximizing the route score and satisfying the travel time constraints. More technical details can be found in [9].

We have evaluated the performance of TripPlanner over the datasets collected from San Francisco. 15,759 POIs of the city were obtained from FourSquare. We ranked all POIs in the descend order based on their total check-in times. The top 1000 POIs were finally used in our work, considering that tourists would seldom visit POIs with few check-ins. The taxi GPS data of San Francisco was obtained from the CRAWDAD[2] data sharing website. Two similar queries with the same time budget (8.5 hours) but different trip starting time were predefined. The start time of the two queries was set to 7:00 am and 10:00 am, respectively. As shown in the right of Fig. 3, given the time budget, the user can have seven preferred venues for the first query while eight for the second one. It is because that, for the first query, the planned route starts around the morning rush hour and thus needs more transit time. We can also find that the route for the two queries are different. These results indicate that TripPlanner is traffic-aware with the usage of taxi GPS data. Moreover, with LBSN data in use, our method is more venue-aware. For example, the user is suggested to go to an Italian Restaurant for lunch just after leaving the Airport, since it is almost lunch time, and many people visit that venue during that time period.

---

[1] http://www.foursquare.com/

[2] http://crawdad.org/index.html

## 4.2. HMI in Crowd-Powered Data Transmission

The success of MCSC relies on the effective transmission of data from individual mobile devices to the destination nodes (e.g., data requesters, backend servers). The mobility of mobile devices and their carriers not only provides a nice coverage for sensing tasks but also brings challenges to data transmission. For instance, both network topology, device connectivity, and communication interfaces evolve over time, which makes it hard to find stable routes for crowdsensed data transmission. We term it as the "transient" networking issue in MCSC.

To address this issue, people often form opportunistic social networks (OPSNs) [4]. Since the source node and destination nodes may never meet in OPSNs, forwarding data packets from its sender to the nodes of interest is often based on the broker-based solution. In this solution, a selected broker node first stores the data from its sender, carries it while in motion, and then forwards it to intermediate or destination nodes. The assumption is that all users are willing to act as brokers. However, this assumption does not always hold: according to sociological theories, socially selfishness is a basic attribute of human beings, and thus the selected brokers may deny requests from other nodes to save their own resources (e.g., storage, power). Therefore, how to motivate people to participate in opportunistic data transmission becomes a crucial challenge of MCSC. We have developed two HMI-enhanced approaches to promote user participation in broker-based data dissemination. In the first approach people are inspired by social/ethical reasons, while in the second one a solid economical model is leveraged. We illustrate them in Fig. 4 and describe them in detail below.

*(1) Hybrid Social Networking (HSN) Model.* The HSN model [10] is inspired by the multi-community involvement and cross-community traversing nature of modern people. For example, at one moment, *Bob* is staying at a place with Internet connection and he can communicate with his online friends; later, he may travel by train with merely opportunistic connection with nearby passengers. We use HSN to indicate the smooth switch and collaboration between online and opportunistic communities, as shown in Fig. 4 (upper).

One of the key issues addressed by HSN is social selfishness. According to [11], people are willing to help their friends. Following this finding, HSN allows the data sender to choose brokers online from their social connections to avoid the selfishness problem. The online approach also reduces the cost on popular node selection (to shorten transmission latency, the ones with high probability to meet more people are selected as brokers), which does not require direct contacts in the real world. The selected brokers will disseminate the information and make match-making to potential interested nodes in opportunistic communities. We compared the performance of HSN with single-community dependent methods (e.g., the pure opportunistic networking method). Experiment results indicate that great performance improvement is obtained when using HSN [10]. This is because that the integration of online communities shortens the broker selection process, and increases the opportunity to select popular brokers.

HSN is a typical HMI-powered approach, where HI and MI are fused in a hybrid manner via two units. The prior unit is sequential, where social relation and user popularity (HI) are derived first and used then for broker selection (MI); the latter one is parallel, where user movement (HI) and match-making (MI) work simultaneously to fulfill the data transmission task.

*(2) Market Model with Intermediaries*. Inspired by how buyers and sellers interact in traditional markets, we introduce the model of markets with intermediaries as an incentive mechanism to stimulate node cooperation in MCSC. In many markets (e.g., stock markets, agricultural markets in developing countries), individual buyers and sellers do not interact directly with each other, but trade via intermediaries instead [11]. These intermediaries, also called traders, often set the prices of transactions. The similarities between markets with intermediaries and data dissemination in opportunistic social networks drive us to use the former as an incentive mechanism in the latter. A data sender is like a seller in a market, and a data receiver is like a buyer, she "buys" a unit of good if she receives the data from traders. As shown in Fig. 4 (bottom), the connections (built based on direct contacting) among traders, sellers, and buyers will form a trading network.

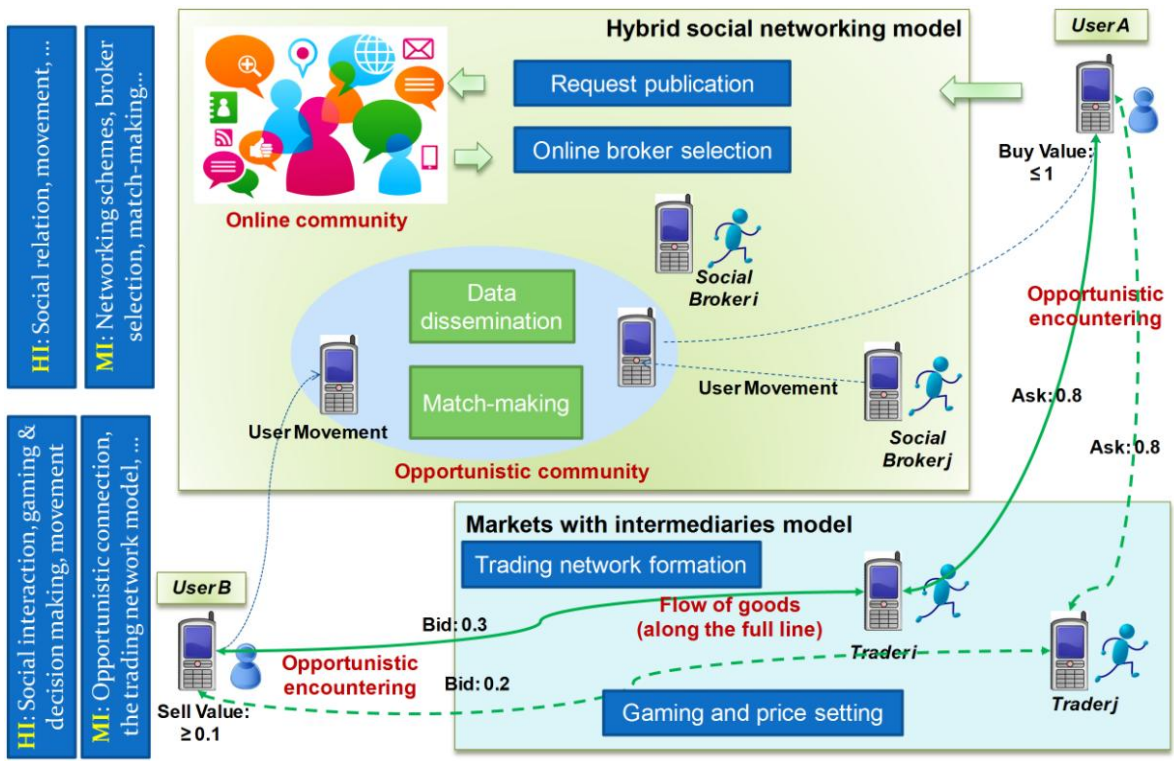

Fig. 4. HMI in two crowd-powered data dissemination models

For simplicity, each seller $i$ initially holds one unit of the good, which she values at $v_i$; she is willing to sell it at any price that is at least $v_i$ (*sell value*). Each buyer $j$ values one copy of the good at $v_j$ (*buy value*) and will try to obtain a copy of the good if she can do it by paying no more than $v_j$. Each trader $t$ offers a *bid price* $b_{ti}$ to each seller $i$ that she connects, and an *ask price* $a_{tj}$ to each connected buyer $j$. After receiving offered prices by traders, each seller and buyer can only choose at most one trader to deal with. A *flow of goods* from sellers, through traders, to buyers is finally generated. Figure 4 gives an example of such a trading model, including the bid price, ask price, and the flow of goods (indicated by the solid lines).

In this approach, transactions are made based on a game process. In the game, a trader's strategy is a choice of bid and ask prices to propose to each neighboring seller and buyer; a seller or buyer's strategy is a choice of a neighboring trader to deal with, or the decision not to take part in a transaction. The participants (sellers, traders, and buyers) are motivated to get their payoffs. Noted that in traditional markets, currency is generally used as a medium for buying and selling. In our model, we further expand this concept by allowing virtual currency to be used. That's to say, services can pay "virtual coins" to the participants and participants need to spend some coins in service usage. In our case, both data senders and brokers can receive their payoffs in "virtual coins". Experiments indicate that our approach can enhance user participation in MCSC data transmission [12]. HMI is also embedded in a hybrid way in this case. In association with social interactions and price settings (HI), the trading network model (MI) is formed (a parallel unit), afterwards the stakeholders bargains and make decisions (HI) to fulfill the data transmission task (a sequential unit).

## 4.3. Other Efforts on MCSC

Beyond our recent works, MCSC is also found useful in several other studies. Zheng et al.[13] had proposed a model to infer fine-grained air quality information throughout a city. The learning model leveraged the air quality data reported by existing monitor stations and a variety of crowd-contributed data in the city, such as traffic flow (offline space) and POIs in LBSNs (online space). Du et al. [14] leveraged a combination of social media and historical physical activity data to predict activity attendance and facilitate social interaction in the real world.

## 5. Conclusion

MCSC shows its difference in the literature history by leveraging various user participation in data contribution and aggregating the heterogeneous crowdsourced data for novel service provision. We have characterized the key features of MCSC, such as explicit/implicit sensing, and heterogeneous, cross-space data mining. To fully leverage the power of crowd participation, MCSC needs the deep fusion of human and machine intelligence. Three HMI patterns are thus identified. We further present the early efforts to MCSC.

As an emerging paradigm for large-scale sensing, numerous challenges and research opportunities remain investigated. First, MCSC is an instance that bridges the gap between the cyber space and the physical space. The problems to solve in such a hyperspace are much more complex and need to integrate various HI and MI units as interdependent parameters in a unique solution. Second, in the hyperspace, we should exploit cross-space features for aggregated sensing and data understanding. Third, as community-enabled sensing, the generic features of a community, such as sensing scale (ranging from a group to the urban scale), community structure, and user collaboration should be further studied [5, 15], which are paid little attention in existing studies.

## Acknowledgement

This work was partially supported by the National Basic Research Program of China (No. 2015CB352400), the National Natural Science Foundation of China (No. 61332005, 61373119, 61222209), and the Scientific and Technology New Star of Shaanxi Province (2014KJXX-39).

# Biographies

BIN GUO (guob@nwpu.edu.cn) is a professor from the School of Computer Science, Northwestern Polytechnical University, China. He received his Ph.D. degree from Keio University, Tokyo, Japan, in 2009. During 2009-2011, he was a post-doctoral researcher at Institut TELECOM SudParis in France. His research interests include pervasive computing, social computing, and mobile crowd sensing. He has served as an editor or guest editor for a number of international journals, such as IEEE Communications Magazine, IEEE Transactions on Human-Machine Systems (THMS), IEEE IT Professional, Personal and Ubiquitous Computing (PUC), and ACM Transactions on Intelligent Systems and Technology (TIST). Dr. Guo has served as the general/program/workshop chair for several conferences, including IEEE UIC, IEEE SCI@PerCom, IEEE iThings, etc.

CHAO CHEN (chunchaaonwpu@gmail.com) is an associate professional from Chongqing University, China. He received his Ph.D. degree from Pierre and Marie Curie University, France in 2014. His research interests include pervasive computing, social network analysis, and big data analytics for smart cities. He is a member of IEEE.

DAQING ZHANG (daqing.zhang@it-sudparis.eu) is a professor from Institut TELECOM SudParis, France. He obtained his Ph.D. from University of Rome "La Sapienza", Italy in 1996. He has organized a dozen of international conferences as General Chair or Program Chair. He is the associate editor for four journals including ACM TIST, Springer Journal of Ambient Intelligence and Humanized Computing, etc. He also served in the technical committee for conferences such as UbiComp, Pervasive, PerCom, etc. Dr. Zhang's research interests include ubiquitous computing, context-aware computing, big data analytics, and social computing.

ZHIWEN YU (zhiwenyu@nwpu.edu.cn) is a professor and director of the Department of Discipline Construction, Northwestern Polytechnical University, China. He has worked as an Alexander Von Humboldt Fellow at Mannheim University, Germany from Nov. 2009 to Oct. 2010, and a research fellow at Kyoto University, Japan from Feb. 2007 to Jan. 2009. His research interests cover pervasive computing, context-aware systems, and personalization. Dr. Yu has served as an editor or guest editor for a number of journals, such as IEEE Communication Magazine, PUC and ACM TIST. He is the General Chair of UIC 2014 and the Workshop Chair of UbiComp 2011.

ALVIN CHIN (alvin.chin@utoronto.ca) is a Senior Researcher at Microsoft (formerly Nokia) and previously was in the Mobile Social Experiences group at Nokia Research Center, Beijing from 2008 to 2012. He has Bachelors and Masters degrees in Computer Engineering from the University of Waterloo and a PhD in Computer Science from the University of Toronto, Canada. His research interests include social networking and ubiquitous computing. He has served as general/program/workshop chairs for several conferences including IEEE CPSCom, IEEE CIT, IEEE UIC, ACM Hypertext, etc. He is an Associate Editor for the New Review of Hypermedia and Multimedia (NRHM) journal and Co-Editor of the Special Issue in Smartphone-based Technologies, Applications and Systems of ACM TOMM journal.